\documentclass[amsmat,amssymb,amsfonts,aps,prb,twocolumn,showpacs]{revtex4}

\usepackage{amsmath}
\usepackage{graphicx}

\usepackage{xcolor}

\begin{document}

\title{Large spin splitting in  the conduction band of transition metal dichalcogenide monolayers}

\author{K. Ko\'smider,
        J. W. Gonz\'alez\footnote{Current Affiliation - Centro de F\'{i}sica de Materiales (CSIC-UPV/EHU)-Material Physics
                                  Center (MPC), Paseo Manuel de Lardiz\'abal 5, 20018, San Sebasti\'an-Spain},
        J. Fern\'andez-Rossier\footnote{On leave from Departamento de F\'isica Aplicada, Universidad de Alicante,  Spain} }
\affiliation{  International Iberian Nanotechnology Laboratory (INL),
Av. Mestre Jos\'e Veiga, 4715-330 Braga, Portugal
}

\begin{abstract}

We study the conduction band spin splitting that arises in   transition metal dichalcogenide (TMD) semiconductor
monolayers such as MoS$_2$, MoSe$_2$, WS$_2$ and WSe$_2$ due to the combination of spin-orbit coupling and lack of
inversion symmetry. Two types of calculation are done. First, density functional theory (DFT)
calculations based on plane waves  that yield large splittings,  between 3 and 30 meV.
Second, we derive a tight-binding  model,  that permits to address the atomic origin of the splitting.  The basis set of
the model is provided by the maximally localized Wannier orbitals, obtained from the DFT calculation, and formed by 11
atomic-like orbitals corresponding to $d$ and $p$ orbitals of the transition metal (W,Mo) and chalcogenide (S,Se)
atoms respectively.
In the resulting Hamiltonian we can independently change the atomic spin-orbit coupling constant of the two atomic
species at the unit cell, which permits to analyse their contribution to the spin splitting at the high symmetry points.
We find that ---in contrast to the  valence band---  both atoms give comparable  contributions to the conduction band splittings.
Given that these materials are most often  $n-$doped, our findings are important for developments in TMD spintronics.
\end{abstract}

\maketitle


\section{Introduction}

Spin-orbit and exchange are the two dominant spin dependent interactions in solids. Whereas the exchange splitting is only present in magnetic materials, spin-orbit coupling (SOC) is ubiquitous.  The proposal of various physical effects driven by spin-orbit interactions, such as the Spin Hall effect, both extrinsic\cite{Hirsch99} and intrinsic,\cite{Murakami03,Sinova04} as well as the Quantum Spin Hall phase,\cite{Kane-Mele05,Bernevig06} together with their experimental confirmation,\cite{Valenzuela06,Konig07} is opening new venues,
 enlarging the set of materials that could give rise to practical spintronic devices.  The effects of spin-orbit interaction are particularly notorious in materials without inversion symmetry,\cite{RMP-Niu} because they present spin splitting of the bands and the anomalous velocity is non-zero.  

From this perspective, the new generation\cite{review1,review2,review3,Geim13} of semiconducting two dimensional (2D) crystals,
such as transition metal dichalgonides (TMD) monolayers (ML), is particularly appealing.  The spin-orbit interaction of
the constituent atoms is large and those 2D crystals have no inversion symmetry. The resulting spin splitting of the bands
gives rise to the so called spin-valley coupling,\cite{Xiao} which has been experimentally confirmed.\cite{review3,SV1,SV2,SV3,SV4,Zeng2012}
This effect is conspicuously apparent in the valence band (VB) of these materials showing
the SOC splittings ranging between  150 meV (MoS$_2$) up to  to 400 meV (WSe$_2$). 
The effect of SOC in the conduction band (CB), in contrast, have been
overlooked except for a few instances.\cite{Kadantseva12,Cheiw12,Kosmider13,Kormanyos13} 
 However, given that very often 2D TMD can be $n$-doped and the  conduction band spin splitting is definitely non-zero, 
 it is of the largest interest to study this effect. 

The origin of the large spin-orbit splitting of the VB is well understood: at the $K$ points, the  valence band Bloch states wave are  mostly made of the metal $d$ orbitals with $\ell=2, m_\ell= 2 \tau$, where $\tau=\pm 1$ labels the valley index.\cite{Zhu12}
Therefore, the $m_\ell S_z$ component of SOC naturally gives a valley dependent splitting of the bands.  In contrast,
the dominant contribution  of the CB lowest energy state  comes from the $\ell=2, m_\ell$=0 orbitals, that
cancels the spin orbit splitting, calculated at first in perturbation theory. 
Thus, the  conduction band spin splitting was  neglected in the influential seminal work  of Xiao {\em et al.}  \cite{Xiao}, proposing a \textbf{k}$\cdot$\textbf{p}  model,  and most of the papers
that followed.  Only very recently attention is being payed to the conduction band splitting,\cite{Ochoa-Roldan, Kormanyos13}
 using an extension of
the  \textbf{k}$\cdot$\textbf{p} original model,\cite{Xiao} showing that inter-band coupling to remote bands,
results in a finite CB splitting. However, the use of Bloch states inherent in this method obscures the  atomic
origin of the spin splitting, which  remains to be determined and is the main focus of the present work.

In order to address the relative contributions to the CB splitting of the two chemical species in the unit cell it
would be convenient to describe the electronic structure of MoS$_2$ and related 2D crystals in terms of localized atomic
orbitals.  However, most of the existing density functional theory (DFT)\cite{KS-PR} calculations of the spin-orbit properties of these materials use plane
waves as a basis set.\cite{Kormanyos13,Kosmider13,Zhu12, Ramasubramaniam11, Ramasubramaniam12, Britnell,Wirtz13}  In order to bridge the gap
between plane wave and atomistic descriptions we make use of the tight-binding (TB) Hamiltonian with a minimal basis set formed by the maximally localized Wannier functions (MLWFs). \cite{Marzari}
This approach keeps a precision of the plane wave calculations,\cite{Shi13,Espejo13} and at the same time  allows a description of the 
spin-orbit coupling  using  the intra-atomic terms $\lambda\hat{\textbf{L}}\cdot\hat{\textbf{S}}$.
Similar approach (we call it TB+SOC) was used to study the Bi$_2$Te$_3$, Bi$_2$Se$_3$, Sb$_2$Te$_3$ topological insulators,\cite{Zhang10} 
and its important advantage is that the atomic SOC of the transition metal (TM) and chalcogenide (CH) atoms, 
can be varied as the $\lambda_{\rm TM}$ and $\lambda_{\rm CH}$ parameters, which permits to trace the origin of the spin splitting of the different bands. 
The study of spin orbit coupling physics in other two dimensional crystals, such as graphene, using the atomic 
$\lambda\hat{\textbf{L}}\cdot\hat{\textbf{S}}$  
 Hamiltonian has revealed very fruitful in the past.\cite{Min06,Huertas06,Huertas09,Neto09,Fabian2010,Dani2011,Fratini13}

The rest of this paper is organized as follows. In section \ref{DFT_calculations} we present the  DFT methodology and the  electronic
structure of the four two dimensional crystals studied here,  MoS$_2$, WS$_2$, MoSe$_2$ and WSe$_2$. In section \ref{MLWF_basis} we describe the way of obtaining the MLWF basis and the resulting
TB Hamiltonian. In section \ref{atomicSOC} we include SOC to the TB Hamiltonian as a sum of atomic terms which depend on
the $\lambda_{\rm TM}$ and $\lambda_{\rm CH}$ parameters, and we determine values of these parameters.  In section \ref{CB_SOC} we take advantage of the model of
section \ref{atomicSOC} to discuss the relative contribution to the spin-orbit splitting of the conduction band of the two chemical species
of the unit cell.  In section \ref{Conclusions} we discuss the limitations of the model and we present our main conclusions.

\section{Electronic structure using DFT}\label{DFT_calculations}

We now review the electronic properties of the MoS$_2$, MoSe$_2$, WS$_2$, and WSe$_2$ MLs calculated with DFT
 in the plane wave basis as implemented in the {\small VASP} package.\cite{VASP1}
 We take an energy cutoff $E_{\rm cut}$=400 eV. We use the projector-augmented waves (PAW)\cite{Blochl1,Kresse1}
method with the $4p$, $5s$, $4d$ valence states of the TM atom,  and the $3s$, $3p$ valence states
of the CH atoms. The  Perdew-Burke-Ernzerhof's\cite{PBE} version of generalized gradient approximation is used to describe
the exchange correlation density functional.  We use the super-cell of the 1$\times$1 periodicity and a vacuum not thinner than 17 {\AA}. 
 The Brillouin Zone (BZ) is sampled with the $\Gamma$-centered ($9\times9\times1$) Monkhorst-Pack's\cite{Monkhorst} mesh of
\textbf{k}-points. We carry out two kinds of calculations. One with SOC included and the second one without SOC.
From now on we refer to them as to the DFT and DFT+SOC respectively. In the DFT+SOC calculations we use non-collinear
version\cite{Hoobs} of the PAW method and SOC is described using the spherical part of the Kohn-Sham potential inside
the PAW spheres.\cite{KresseSOC}

\begin{figure}[t]
    \centering
    \includegraphics[clip=true, width=0.48\textwidth]{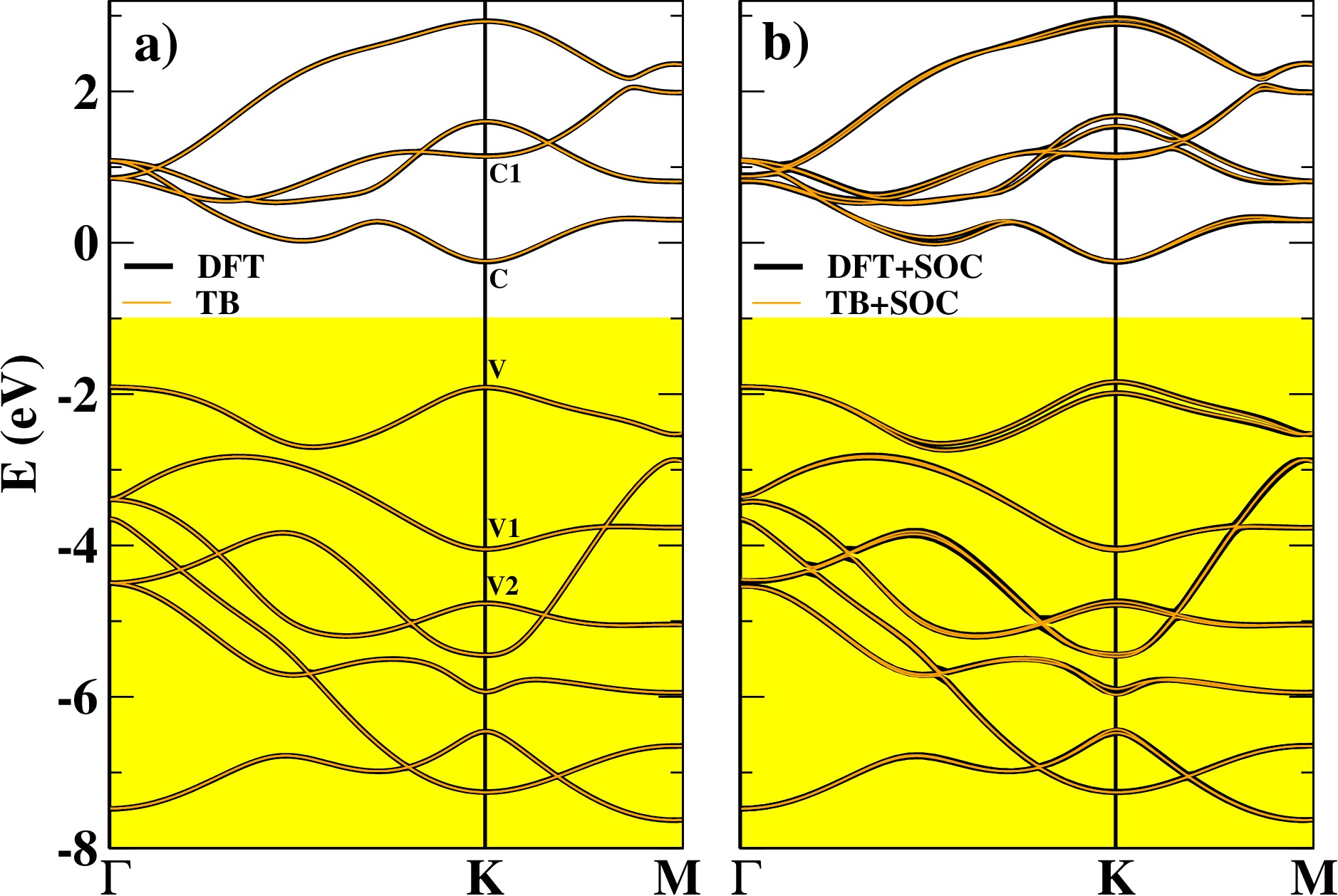}
    \caption{\label{bands}(Color online) Electrons energy band structures of the MoS$_2$ ML calculated with the DFT and
                    TB methods: a) bands calculated without SOC, b) bands calculated with SOC.}
\end{figure}

\begin{table}
\centering
\begin{tabular}{llrrrr}
\hline\hline
& & \multicolumn{4}{c}{$\Delta_n$ (meV)}  \\
\cline{3-6}
  Method& $n$ & \multicolumn{1}{c}{MoS$_2$} & \multicolumn{1}{c}{WS$_2$} & \multicolumn{1}{c}{MoSe$_2$}  & \multicolumn{1}{c}{WSe$_2$}  \\
\hline
DFT+SOC       & C      & \textbf{-3}   &  \textbf{27} &  \textbf{-21} &  \textbf{38} \\
              & V      & 147 & 433 & 186 & 463 \\
              & V1     & 24  &  70 &  27 &  88 \\
              & V2     &-50  & -55 &-188 & -232 \\
\hline
TB+SOC        & C       &    \textbf{-4} &  \textbf{17} &  \textbf{-28} &  \textbf{-3} \\
              & V       &  147 & 433 & 186 & 463 \\
              & V1      &   24 &  66 &  29 &  64 \\
              & V2      &  -50 & -55 &-188 & -232 \\
\hline
PT            & C       &    \textbf{-1} &  \textbf{13} &  \textbf{-11} &  \textbf{7} \\
\hline
\hline
\end{tabular}
\caption{\label{TAB1} The spin-orbit splittings $\Delta_n$ ($n$ labels bands, see Fig.\ref{bands}) at the $K$ point 
calculated for the considered TMD MLs with the DFT+SOC, TB+SOC, or perturbation theory (PT) method.}
\end{table}

We briefly summarize the main features of the MoS$_2$ ML energy bands, as given by DFT and DFT+SOC (Fig.\ref{bands}(a) and 
Fig.\ref{bands}(b) respectively). The results for the MoSe$_2$, WS$_2$ and WSe$_2$ MLs are very similar, and agree with
the previous calculations.\cite{Zhu12,Kadantseva12,Cheiw12,Ramasubramaniam12,Kosmider13} 
The band gap of these semiconducting MLs is direct,  with the minimum of the CB and the top of the VB located at the $K$ and $K'$ points of the BZ.
All the bands at the $K$ points are spin split, but only in some instances the splitting is so large that is appreciated by
inspection in Fig.\ref{bands}(b).

Analysis of the wave functions reveals that it is possible to assign a spin projection along the normal to the plane to
the different Bloch states in the neighbourhood of the  $K$ point. Taking advantage of this, in the following we define
the splitting of a energy band $n$  with momentum $\textbf{k}$ as
\begin{equation}
\Delta_n(\textbf{k}) \equiv \epsilon_{n\uparrow}(\textbf{k}) - \epsilon_{n\downarrow}(\textbf{k}).
\end{equation}
With this convention, the splitting can be either positive or negative. Time reversal symmetry warrants that
$\Delta_n(\textbf{k})=-\Delta_n(-\textbf{k})$ which implies that spin splittings have opposite  signs in $K$ and $K'$ valleys.\cite{Xiao}
The spin splittings of the relevant bands at the $K$ point  are listed in Table \ref{TAB1}.
The spin-orbit splitting at the top of the VB, range between 147 meV for MoS$_2$ and 463 meV for WSe$_2$. 
The same splittings for the CB vary from $-3$ meV for MoS$_2$ to 38 meV for WSe$_2$. They are smaller than those of the VB,
but definitely large enough as to be observed.
 It is worth noticing that only in the case of the conduction band the sign of $\Delta$ is not the same for all the compounds, for reasons explained  below.

We now discuss the population analysis of the DFT Bloch states. This  sheds some light on the origin of their spin
splittings. Both VB and CB are predominantly
made of the TM atom $d$ ($\ell$=2) orbitals and a smaller but not negligible contribution coming from the
$p$ ($\ell$=1, $m_\ell$=$\pm1$) orbitals of the  chalcogen atoms. The main difference between VB and CB bands lies
in the $m_\ell$ number of $d$ orbitals, which is equal $\pm2$ in the VB and 0 in the CB. 
This picture is in agreement with earlier work.\cite{Xiao,Kormanyos13,Kosmider13,Cheiw12,Kadantseva12,Zhu12}
In the  discussion below we shall also make use the fact that the Bloch state labelled as V2 at the $K$ point is made exclusively of the chalcogen $p$ orbitals ($\ell$=1, $m_\ell$=+1), without mixing to the metal $d$ orbitals.

\section{Maximally localized Wannier functions basis} \label{MLWF_basis}

The Wannier functions \cite{Wannier} (WF) permit to  define a localized basis set by performing a unitary transformation
over a set of  Bloch states that diagonalize the DFT Hamiltonian. Although there is not a unique way of doing such a
$wannierization$,  we adopt the method criteria of  maximal localization\cite{Marzari} and 
we use the \texttt{Wannier90}\cite{WANNIER90} code to find the basis of MLWFs.  
This approach has already been used  for MoS$_2$ and related transition metal dichalcogenides before,\cite{Feng_PRB2012} obtaining results in line with those
discussed here. 
In our case, the set is formed by the group of  11 bands distributed around the band gap, as shown in Fig. \ref{bands}(a).

The  first step of the procedure consist of the  projection of the the Bloch states $|\psi_{\small \textbf{k},n}\rangle$ over certain a set of localized functions which, in this case, are taken as the $p$ and $d$ atomic orbitals of the chalcogenide and metallic atom respectively,   motivated by  the population analysis discussed above.
Importantly,  in the case of 2D TMDC, the MLWF are centered around the atoms, their localization radius
 is smaller than the interatomic distance and, in the neighborhood of the atoms, 
 they have the symmetry of the real spherical harmonics.   A numerical measure of the localization is given by
 the  localization functional\cite{Marzari} $\Omega$. In our case,  after 100 iterative steps,  we obtain a  total spread
18.23/20.85/20.28/23.25 {\AA}$^2$, summing over the 11 Wannier orbitals,  for  MoS$_2$/MoSe$_2$/WS$_2$/WSe$_2$, which yields an average size per Wannier orbital of 1.29/ 1.38/1.36/1.45 {\AA}.

\begin{figure}[hbt]
    \centering
    \includegraphics[clip,width=0.48\textwidth,angle=0]{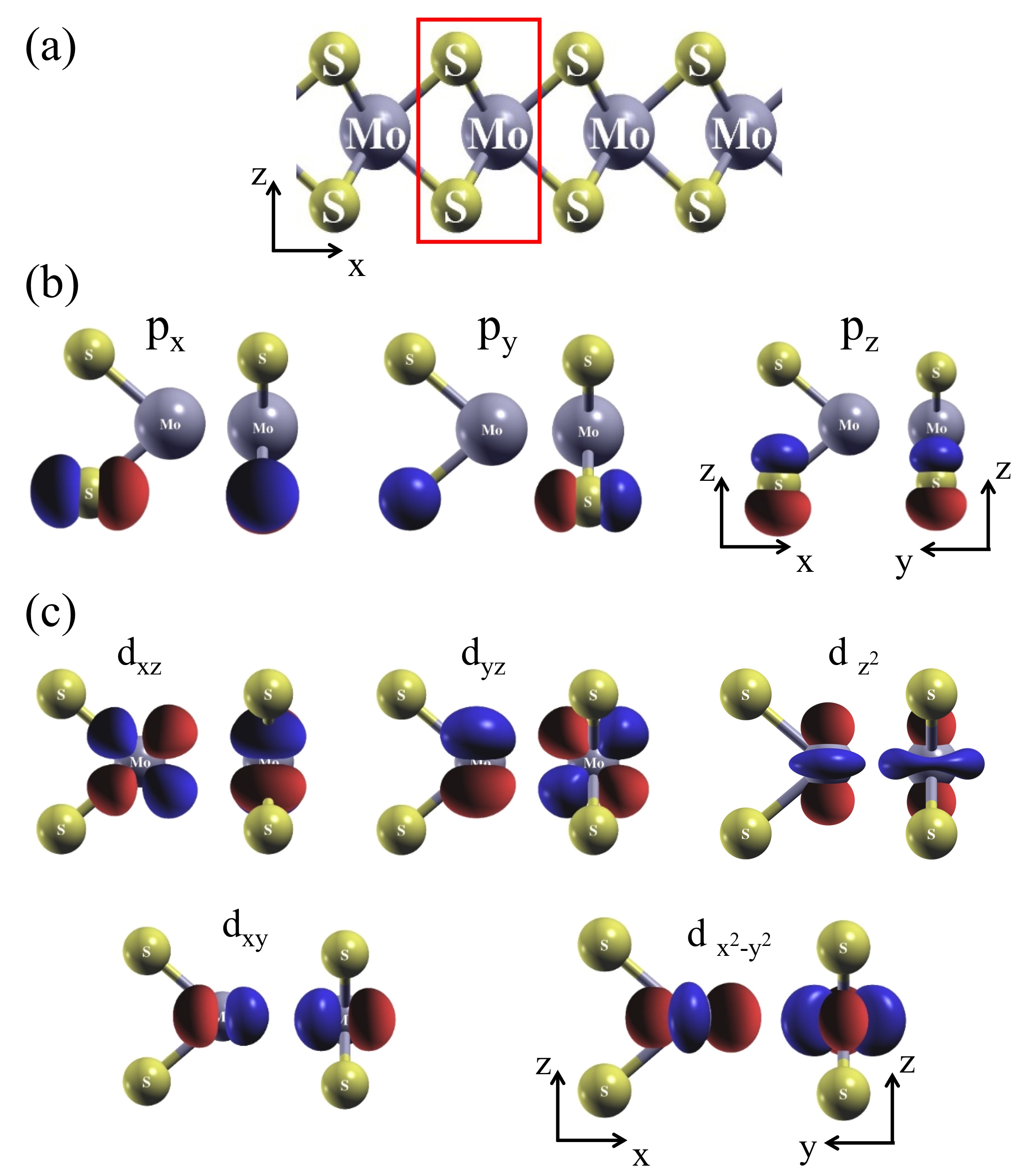}
    \caption{\label{FIG2}(Color online) The MLWF basis used to express TB Hamiltonian (\ref{ham_W}) of the MoS$_2$ ML: (a) Side view of the monolayer, (b) and (c) contour-surface plot
of the three $p$ orbitals of sulphur and the five $d$ orbitals of molybdenum. Figure prepared with XCrySDen.\cite{XcrysDen}}
\end{figure}

The isosurfaces of the MLWF  obtained for MoS$_2$  are presented in  Fig. \ref{FIG2}. Their real spherical harmonic 
symmetry is apparent. In the following we label  the MLWF as $|\textbf{R}O\rangle$, where $\textbf{R}$ defines a unit
cell inside the crystal and $O$ refers to  one  the 11 atomic-like MLWF inside the unit cell.  We refer to them using 
their real spherical harmonic symmetry, as shown in Fig. \ref{FIG2}. However, the shape (not shown) of the tails of the
MLWF  is  different from that of the core.  Therefore, MLWFs   are not  identical to atomic orbitals for which the angular
symmetry is independent of the distance to the nuclei.

\subsection{Wannier Hamiltonian}
The wannierization procedure  yields the basis of 11 atomic-like orbitals $|\textbf{R}O\rangle$, and ---more importantly--- a faithful 
representation of the DFT Hamiltonian in that basis.  Thus, for a given pair of the atomic-like MLWF orbitals $O$ and
$O'$,  located in unit cells $\textbf{R}$ and $\textbf{R}'$, we obtain the representation of the DFT  Hamiltonian $\langle \textbf{R}O|H|\textbf{R}'O'\rangle$. 
  Taking advantage of the Bloch theorem, the Hamiltonian for the entire crystal can be  block diagonalized in the usual way, resulting in the following wave-vector dependent Hamiltonian matrix: 
\begin{equation}\label{ham_W}
 {\cal H}_{OO'}(\textbf{k}) = \sum_{\textbf{R}} e^{i\textbf{k}\cdot \textbf{R}}
  \langle \textbf{0}O|H_{\rm DFT}|\textbf{R}O'\rangle,
\end{equation}
where the sum runs over all the unit cells of the crystal, labelled with $\textbf{R}$. In practice, the localized nature
of the MLWFs permits to truncate the sum down to a few neighbors. Importantly, the dimension of
the matrix  (\ref{ham_W}) is as small as the size of the MLWF basis (11 in the present case) which makes the numeric diagonalization computationally inexpensive.  The
  resulting energy bands are ---not surprisingly (given their formal equivalence)---  very similar to those obtained from DFT as shown in Fig. 
\ref{bands}(a). Minor differences (not appreciated at the energy scale used in the figure) arise from the truncation in 
the number of bands, i.e., due to inter band coupling to remote high and low energy bands that have been excluded in the
Wannier Hamiltonian but are present in the DFT calculation.

The eigenstates of Hamiltonian (\ref{ham_W}) are a linear combination of the MLWFs which ---as discussed above---  have real spherical harmonic symmetry close to the atom cores. In order
to understand the spin splittings, it is convenient to define a new basis of orbitals localized around atom $A$, denoted by
 $|A^{\ell}_{m_\ell}\rangle$, 
 which has the symmetry of the eigenstates of the atomic angular momentum operator.  In other words, we move from a real basis to the usual spherical harmonics with well defined $m_\ell$. 
 In the rest of this paper we use the following notation to relate the Bloch states at the $K$ point
 with the atomically localized orbitals $|A^{\ell}_{m_\ell}\rangle$:
 \begin{equation}
 \label{wfunc}
  |\psi_{K,n}\rangle =  \alpha_n |{\rm TM}^{\ell=2}_{m_\ell}\rangle + \beta_{n}
   \left(|{\rm CH1}^{\ell=1}_{m_\ell}\rangle + s_n |{\rm CH2}^{\ell=1}_{m_\ell}\rangle \right),
\end{equation}
where  $\alpha_n$ and $\beta_n$ are coefficients, and $s_n=\pm1$  ($+1$ for the bands C, V, V1 and $-1$ for C1 and V2).

\begin{table}[h]
\centering
\begin{tabular}{lrrcccccccccccc}
\hline\hline
              & \multicolumn{2}{c}{$m_\ell$} && \multicolumn{2}{c}{MoS$_2$}   && \multicolumn{2}{c}{WS$_2$} && \multicolumn{2}{c}{MoSe$_2$} && \multicolumn{2}{c}{WSe$_2$} \\
\cline{2-3} \cline{5-6} \cline{8-9} \cline{11-12} \cline{14-15}
          $n$  &  $|$TM$\rangle$ & $|$CH$\rangle$ && $\alpha_n^2$ & $\beta_n^2$ && $\alpha_n^2$ & $\beta_n^2$ && $\alpha_n^2$ & $\beta_n^2$ &&  $\alpha_n^2$ & $\beta_n^2$  \\
\hline
C1        &  $-1$ &$+1$ && 0.63 & 0.19 & &  0.63 & 0.19 && 0.65 & 0.18 && 0.65 & 0.18 \\
C         &   0   &$-1$ && 0.86 & 0.07 & &  0.90 & 0.05 && 0.86 & 0.07 && 0.89 & 0.05 \\
V         &  $+2$ &$+1$ && 0.80 & 0.10 & &  0.79 & 0.11 && 0.82 & 0.09 && 0.79 & 0.10 \\
V1        &  $+1$ &$0 $ && 0.28 & 0.36 & &  0.25 & 0.38 && 0.34 & 0.33 && 0.30 & 0.35 \\
V2        &  ---  &$-1$ && ---  & 0.5  & &  ---  & 0.5  && ---  & 0.5  && ---  & 0.5  \\
\hline\hline
\end{tabular}
\caption{\label{TAB2} Table of projections of Bloch states at $K$ over the $|A^{\ell}_{m_\ell}\rangle$ basis
(see Eq. (\ref{wfunc})). The leftmost columns denote the $m_{\ell}$  relevant for each band.
Since there are two equivalent chalcogen atoms per unit cell, the normalization criteria is $|\alpha_n|^2+2 |\beta_n|^2=1$}
\end{table}

Importantly, since the MLWFs do  not  rigorously have  spherical harmonic symmetry, 
the $|A^{\ell}_{m_\ell}\rangle$ are not rigorously eigenstates of the atomic angular momentum operator.
However, in the rest of this work, 
we adopt the approximation that the $|A^{\ell}_{m_\ell}\rangle$ are indeed eigenstates of the 
 atomic orbital angular momentum operator. The validity of this approach is justified by the  fairly good agreement  with the  DFT results, discussed below.

In Table  \ref{TAB2} we show    $|\alpha_n|^2$ and $|\beta_n|^2$.  It is apparent that the CB and
VB are mostly made of the transition metal $d$ orbitals, with $m_\ell$ equal 0 and 2 respectively. The small variations of
the coefficient squares $\alpha^2$ and $\beta^2$ along the different materials inform of their similar electronic
structure. It must be noticed that the contributions of the orbitals localized on the CH atoms is larger than
10$\%$, and thereby they can account for a fraction of the spin splitting, as it actually happens.
Inspection of the wave functions also reveals their odd/even character  with respect to reflection across the $z=0$ plane. Specifically,
 the wave functions of bands C and V are even and those of  bands C1, V1, and V2 are odd, 
  in agreement with  previous results.\cite{Kormanyos13}

\section{Atomic SOC}\label{atomicSOC}
    The  Wannier Hamiltonian just described is derived from a DFT calculation where SOC has been deliberately excluded. We now proceed to add  the atomic spin orbit coupling into the TB Hamiltonian
\begin{equation}\label{ham_SOC}
{\cal \hat{V}}_{\rm SOC}
            = 
           \sum_A \lambda_{_A} \hat{\textbf{L}}_{_A} \cdot \hat{\textbf{S}},
\end{equation}
  where $\lambda_A$ is a scalar that measure the strength of the atomic SOC, $\hat{\textbf{L}}_A$ is the angular momentum operator acting on an atom $A$, and $\hat{\textbf{S}}$ are the spin $1/2$ Pauli matrices operators.  As  discussed after Eq. (\ref{wfunc}),  we assume that
 
  \begin{eqnarray}
&& \langle A^{\ell}_{m_\ell}|\hat{L}^{\pm}_{_A}|A^{\ell}_{m_\ell'}\rangle
  = \sqrt{\ell(\ell +1) - m_\ell(m_\ell' \pm 1)}\delta_{m_\ell,m_\ell '+1} \nonumber, \\
 && \langle A^{\ell}_{m_\ell}|\hat{L}^z_{_A}|A^{\ell}_{m_\ell'}\rangle
  =m_\ell\delta_{m_\ell,m_\ell '}. 
  \end{eqnarray}

The  addition of ${\cal V}_{\rm SOC}$  to  Hamiltonian (\ref{ham_W})  leads to the following TB Hamiltonian
\begin{equation}\label{ham_SOC2}
{\cal \hat{H}}(\textbf{k}) = {\cal \hat{H}}_0(\textbf{k}) + {\cal \hat{V}}_{\rm SOC} ,
\end{equation}
which is the main result of this work. The presence of ${\cal \hat{V}}_{\rm SOC}$ in Eq. (\ref{ham_SOC2}) causes  spin splittings $\Delta_n$, which depend on two parameters $\lambda_{\rm CH}$ and $\lambda_{\rm TM}$.

We are now in position of achieving two goals. First, we can verify the validity of our approach  fitting the $\lambda$  parameters that give a best agreement between the bands of Hamiltonian (\ref{ham_SOC2}) and those obtained with the DFT+SOC method, paying  special attention to the spin splittings 
 $\Delta_n$ close to the $K$ point.  Second, we can determine the  contribution each atom  to the  spin-orbit splitting a various bands, with an attention to the conduction band. 

\subsection{Perturbative estimate of  $\lambda$}
It is very instructive to obtain formal expressions for the $\Delta_n$ splittings treating
${\cal \hat{V}}_{\rm SOC}$ to first order in perturbation theory. A comparison  of these  expressions  with
the values calculated using DFT+SOC method yields a first estimate for $\lambda_{\rm CH}$ and $\lambda_{\rm TM}$.
Choosing  $\hat{z}$ as the spin quantization axis, the shift of the levels with spin $\sigma$, to first order in perturbation theory, reads:
\begin{equation}
\delta \epsilon_{n\sigma}(\mathbf{k}) = \frac{\sigma}{2} \langle
  \psi_{{n}\mathbf{k}}|\sum_A\lambda_{_A}^{} \hat{L}^z_{_A}
   |\psi_{{n}\mathbf{k}}\rangle.
\end{equation}

Since there are two unknowns,  we implement this procedure with two bands,  $|\psi_{_{{\rm V2},K}}\rangle$ and $|\psi_{_{{\rm V},K}}\rangle$  at the $K$ point.  In the case of V2 the contribution from the TM is strictly null, so that first order perturbation theory yields:
\begin{equation}\label{lambdas}
 \Delta_{_{\rm V2}} = \langle
  \psi_{_{{\rm V2},K}}|\sum_A\lambda_{_A}^{} \hat{L}^z_{_A}
   |\psi_{_{{\rm V2},K}}\rangle= -\lambda_{_{\rm CH}} .
\end{equation}
which permits to relate directly the splitting of the V2 band at the $K$ point with the chalcogenide spin orbit coupling. 
In the case of the VB the first order perturbation theory yields:
\begin{equation}\label{lambdas}
 \Delta_{_{\rm V}}   =
2\alpha_{_{\rm C}}^2 \cdot \lambda_{_{\rm TM}}^{} + 2\beta_{_{\rm C}}^2\cdot\lambda_{_{\rm CH}}^{}.\\
\end{equation}
Combining these two equations, we obtain an estimate for $\lambda_{\rm TM}$ and $\lambda_{\rm CH}$, shown 
in the PT columns of Table \ref{tab_lambdas}, together with the estimates using a non-perturbative fitting described below.

The first point to notice is that across different materials (except in the case of Se) the values of $\lambda$ undergo variations smaller than $10\%$.  
This is in line with the general notion that for a given atom,  spin-orbit coupling does not vary much from compound to compound.   These small variations are a first indication of the validity of our  methodology. The second point is that these values are in line with those  reported for neutral S/Se atoms (50/220 meV) \cite{Wittel} as well as for Mo (78 meV  ). \cite{Dunn}  Moreover, it must be kept in mind that the localization of Wannier and atomic orbitals can be different.  Thereby, a scaling of the $\lambda$ for the Wannier orbitals, compared to the atomic orbitals,  is expected.  Our calculations indicate that this is not a large effect, endorsing the notion that the MLWF used in our calculation are similar to the atomic orbitals.

\begin{table}
\centering
\begin{tabular}{lccccc}
\hline\hline
  &  \multicolumn{2}{c}{$\lambda_{\rm TM}$ (meV)} & & \multicolumn{2}{c}{$\lambda_{\rm CH}$ (meV)} \\
\cline{2-3}\cline{5-6}
  &  PT & TB+SOC &  &  PT & TB+SOC \\
\hline
 MoS$_2$           &  87  &  86 & &   50  & 52 \\
 WS$_2$            &  274 & 271 & &   55  & 57 \\
 MoSe$_2$          &  94  &  89 & &   188 & 256\\
 WSe$_2$           &  261 & 251 & &   232 & 439\\
\hline
\hline
\end{tabular}
\caption{\label{tab_lambdas} The atomic SOC parameters $\lambda_{\rm TM}$ and $\lambda_{\rm CH}$ of the considered TMD MLs.
Comparison of the values estimated with perturbation theory (PT) with the values calculated by fitting the $\Delta_{\rm V}$
and $\Delta_{\rm V2}$ splittings to the values obtained from DFT+SOC method (see Table \ref{TAB1}).}
\end{table}

\subsection{Non perturbative determination of $\lambda$}
We now discuss a second and more accurate way to determine the $\lambda_{\rm TM}$ and $\lambda_{\rm CH}$ parameters.  For a given value of   $\lambda_{\rm TM}$ and $\lambda_{\rm CH}$, numerical diagonalization of this Hamiltonian yields a set of spin split bands.

As in the perturbative case, we determine $\lambda$'s   
by fitting the spin splitting at the $K$ point of  both valence and V2 bands to those obtained in the DFT+SOC calculations.
 The values of $\lambda_{\rm TM}$ and $\lambda_{\rm CH}$ parameters estimated this way are listed in the TB+SOC columns of Table \ref{tab_lambdas}.
They are close to the PT values except the $\lambda_{\rm Se}$ in the WSe$_2$ ML.  Possible explanations for this are detailed  below.
\begin{figure}[h!]
    \centering
    \includegraphics[width=0.48\textwidth]{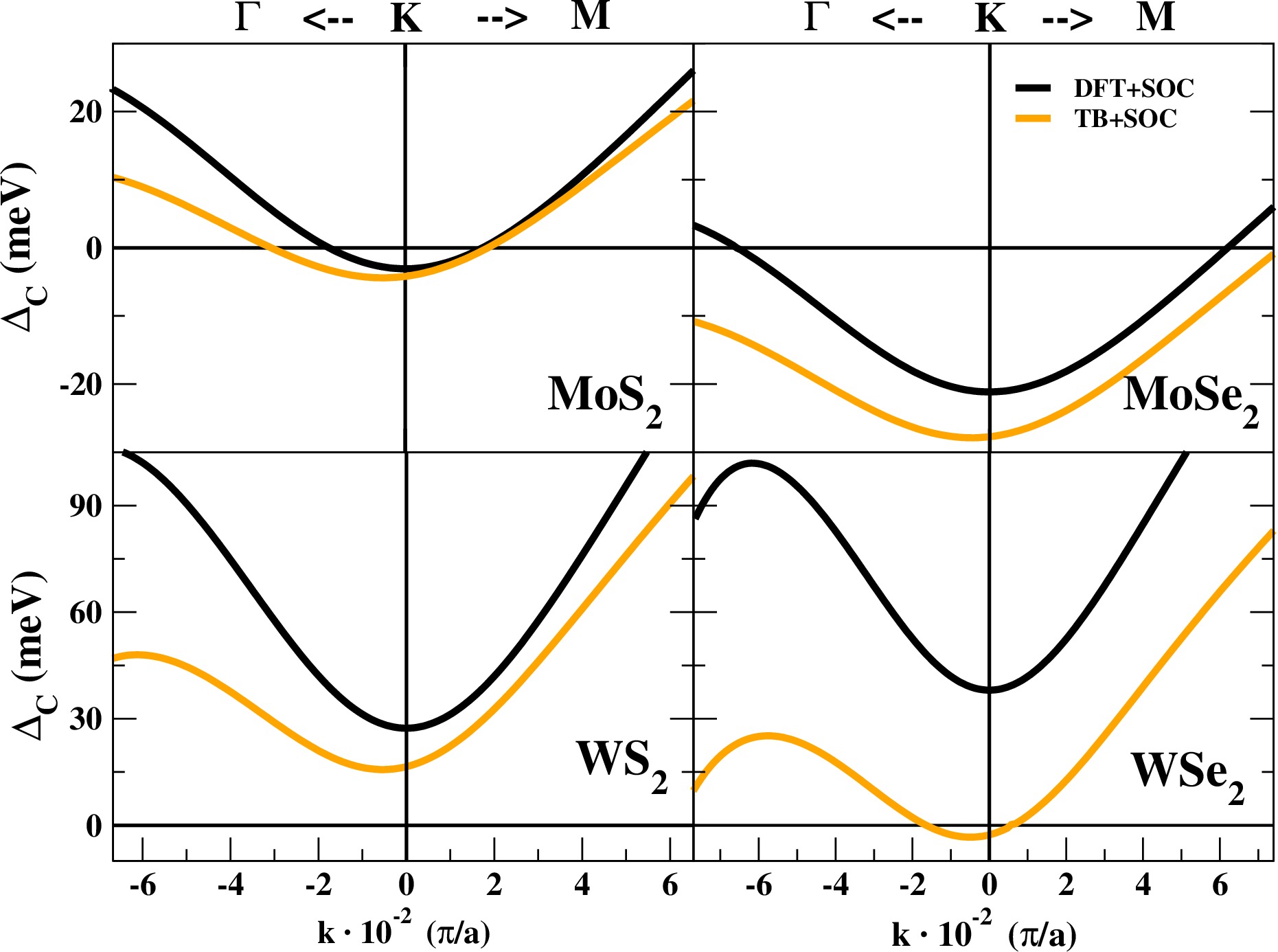}
    \caption{\label{FIG4}(Color online) Comparison of spin-orbit splitting of the conduction band around the $K$ point calculated with the TB+SOC and DFT+SOC methods.}
\end{figure}

In Fig. \ref{bands} we show a comparison of the DFT+SOC bands (left panel) and with the just described TB+SOC method (right panel). It is notorious that,  fixing the splitting of two bands at the $K$ point,  yields a fairly good agreement for all the bands on the entire 
Brillouin zone,  giving additional support to the methodology.

A more quantitative comparison between the TB+SOC and the DFT+SOC calculations is shown in Table
\ref{TAB1} where we  compare the spin splitting of several bands at the $K$ point obtained with the two methods.  Of course, by construction of the method, the agreement for the VB and V2 is perfect.  In addition, it is apparent that the TB+SOC provides a fairly good quantitative agreement for the spin splitting of the conduction and V1 bands, except for WSe$_2$. In Fig. \ref{FIG4} we compare $\Delta_{\rm C}(\textbf{k})$ for DFT+SOC and TB+SOC
 along the $\Gamma-K-M$ high symmetry points.  It is apparent that the TB method captures the non-trivial momentum dependence featured by the DFT+SOC, although there is a systematic off-set which is also  larger for WSe$_2$.

\section{Conduction band spin-orbit splitting}\label{CB_SOC}
\begin{figure}[hbt]
    \centering
    \includegraphics[width=0.48\textwidth]{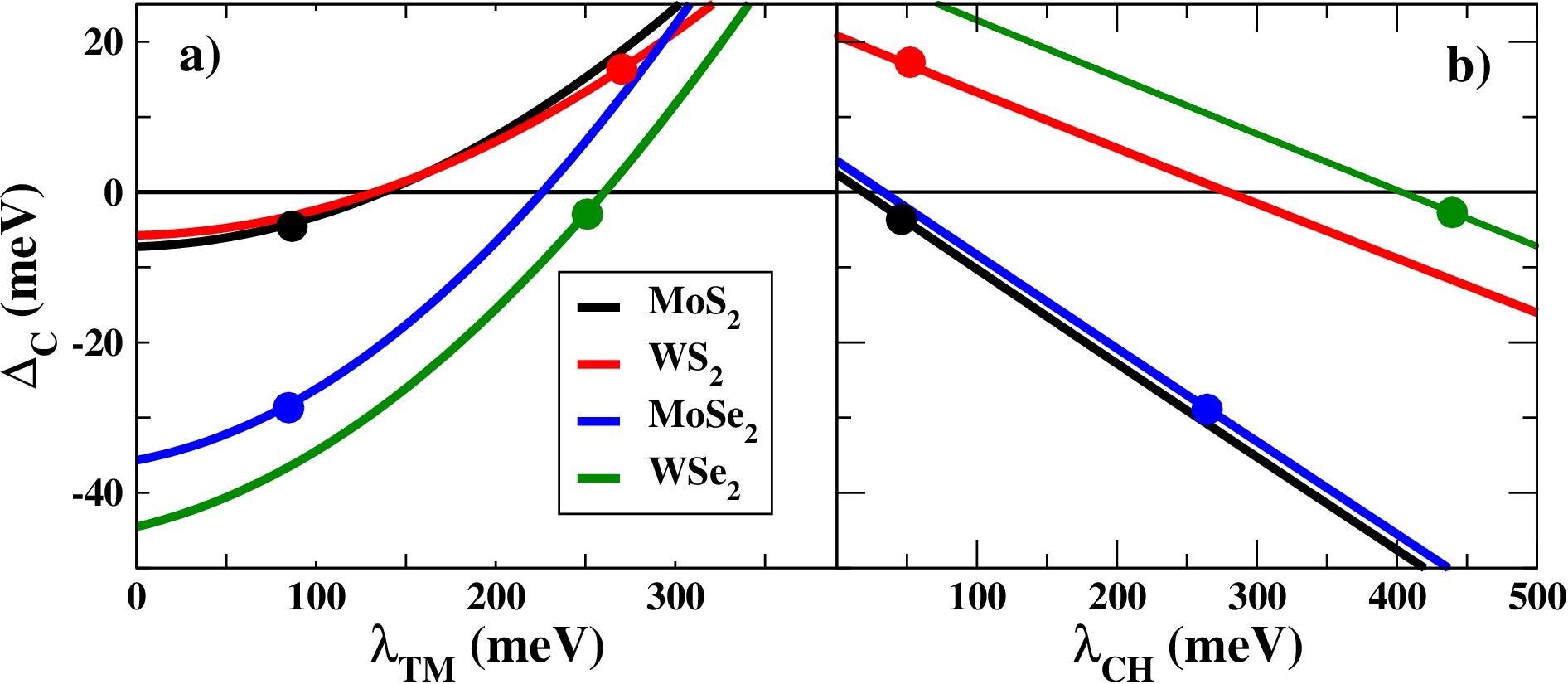}
    \caption{\label{FIG3}(Color online) Spin-orbit splitting of the conduction band at the $K$ point calculate for the considered TMD MLs as a function of $\lambda_{\rm TM}$ (a) and $\lambda_{\rm CH}$ (b) parameters in the. The corresponding $\lambda_{\rm CH}$ (a) and $\lambda_{\rm TM}$ (b) parameters are fixed to the values estimated with the TB+SOC method (see Table \ref{TAB1}). Bold dots mark the $\lambda_{\rm TM}$ and $\lambda_{\rm CH}$ parameters calculated with the same method.}
\end{figure}

We are now in a position to discuss the mechanism for the conduction band splitting in the  TMD monolayers.
Inspection of the $m_\ell$ values in Table \ref{TAB2} reveals, that $\Delta_{\rm C}$ should vanish to first order in $\lambda_{\rm TM}$,
and have a small linear contribution in $\lambda_{\rm CH}$. To check this out we plot in Fig. \ref{FIG3} the $\Delta_{\rm C}(K)$ splitting, keeping one of the $\lambda$ values as given in table III (TB+SOC values), and varying the other. The $\Delta_{\rm C}(\lambda_{\rm CH})$ dependence (for $\lambda_{\rm TM}={\rm const}$) is a straight line with negative slope. This can be understood within 1st order perturbation theory, that yields the following expression for the chalcogen atom SOC contribution to the splitting:
\begin{eqnarray}\label{1st-CB}
\delta \epsilon^{^{\rm(1)}}_{_{\rm C\uparrow}}(K)-\delta \epsilon^{^{\rm(1)}}_{_{\rm C\downarrow}}(K) &=&  \langle
  \psi_{_{K,\rm C}}|\lambda_{_{\rm CH}}^{} \hat{L}^z_{_{\rm CH}}
   |\psi_{_{K,\rm C}}\rangle =\nonumber
   \\&=&  -\lambda_{_{\rm CH}}^{}\beta_{_{\rm C}}^2,
\end{eqnarray}
where the negative sign comes from the fact that, at the $K$ point, the CB Bloch state overlaps with the 
 $m_{\ell}=-1$ chalcogenide atomic like state (see Table \ref{TAB2}).
   With this equation the negative slopes $\partial \Delta_{\rm C}/\partial \lambda_{\rm CH}$ in Fig. \ref{FIG3}(b) became clear.
They are controlled by $\beta_{\rm C}$ (see Table \ref{TAB2}), and are the same for the tungsten based WS$_2$ and WSe$_2$
compounds as well as molybdenum based MoS$_2$ and MoSe$_2$ compounds.
   
  In contrast,  the $\Delta_{\rm C}(\lambda_{\rm TM})$ dependence (for $\lambda_{_{\rm CH}}={\rm const}$) is not linear ---reflecting the inter band character of this contribution--- and has a positive sign.   Of course, opposite signs and trends are found at the  $K'$ point,  on account of time reversal symmetry. The well defined sign of the inter-band contribution to the CB spin splitting is understood as follows. First, we use second order perturbation theory, that  yields  positive (negative) shifts via inter-band coupling to states below (above) in energy.  Second, given the fact that at the $K$ point the Bloch states overlap with states with a well defined handedness, together with the angular momentum conservation, result in a spin-selective inter band coupling.  Thus, the  TM SOC can connect the CB state ($m_\ell=-1$) with spin $\uparrow$ only to the states with the opposite values of $m_{\ell}$ ($+$1) and spin ($\downarrow$), which happen to be available at the band V1, providing a positive contribution of the shift given by 
\begin{equation}\label{second-order-CV1}
\delta \epsilon^{^{\rm(2)}}_{_{\rm C\uparrow}}(K)=\frac{1}{4}
\frac{\lambda_{_{\rm TM}}^2 |\langle\psi_{_{{\rm C},K}}| \hat{L}^{-}|\psi_{_{{\rm V1},K}}\rangle
\langle\uparrow|\hat{S}^{+}|\downarrow\rangle\rangle|^2}{\epsilon_{_{\rm C}}(K)-
\epsilon_{_{\rm V1}}(K)},
\end{equation}
whereas the coupling of the CB to V1 give a null shift of 
$ \delta \epsilon^{\rm(2)}_{\rm C\downarrow}(K)$.
 In contrast,  the CB state with spin $\downarrow$,  can only connect to states with $m_{\ell}=-1,\uparrow$,   which happen to be available at the C1 state,  giving a negative shift to the $\downarrow$ level and thereby another positive contribution to the splitting:
\begin{equation}\label{second-orderCC1}
\delta \epsilon^{^{\rm(2)}}_{_{\rm C\downarrow}}(K)=\frac{1}{4}
\frac{\lambda_{_{\rm TM}}^2 |\langle\psi_{_{{\rm C},K}}|\hat{L}^{+}|\psi_{_{{\rm C1},K}}\rangle
\langle\downarrow|\hat{S}^{-}|\uparrow\rangle\rangle|^2}{\epsilon_{_{\rm C}}(K)-
\epsilon_{_{\rm C1}}(K)},
\end{equation}

We now define $\delta_{\rm C,V1} \equiv \epsilon_{\rm C}(K)-\epsilon_{\rm V1}(K)$ and  
$\delta_{\rm C,C1} \equiv \epsilon_{\rm C}(K)-\epsilon_{\rm C1}(K)$.
Combining Eq. (\ref{1st-CB})-(\ref{second-orderCC1}) with Eq. (\ref{wfunc}),  using $|\langle\uparrow|\hat{S}^{+}|\downarrow\rangle|^2=1$ and 
 $|\langle \psi_{{\rm C},K}|\hat{L}^{-}|\psi_{n,K}\rangle|^2=2\alpha_{\rm C}^2 \alpha_{n}^2  $, with $n$=C1,V1, 
   we can write the following perturbative expression for the CB spin splitting at the $K$ valley: 
\begin{equation}\label{delta_c}
\Delta_{\rm C}(K) = -\lambda_{_{\rm CH}}\beta_{_{\rm C}}^2 + 
\frac{\left(\lambda_{_{\rm TM}}\alpha_{_{\rm C}}\right)^2}{2}
\left(\frac{\alpha_{_{\rm V1}}^2}{\delta_{_{\rm C,V1}}} -\frac{\alpha_{_{\rm C1}}^2}{\delta_{_{\rm C,C1}}} \right).
\end{equation}
However, since $\delta_{\rm C,V1}> 0$ eV and $\delta_{\rm C,C1}<0$ eV,  the two terms proportional to $\lambda_{\rm TM}^2$ are positive.
The  values of $\Delta_{\rm C}(K)$ calculated with Eq. (\ref{delta_c}) for the four  TMD MLs are listed in Table \ref{TAB1} (see row PT).
It is apparent that the perturbative  calculation captures the trend of the non-perturbative results calculation result  and provides a qualitative insight of the
contribution of each atom to the conduction band splitting.

In summary,  the CB splitting has two contributions with opposite signs.
  For the K valley, the chalcogen SOC gives a negative contribution and the transition metal a positive one.  This explains the material dependent sign. Thus, WS$_2$ combines the largest positive with the smallest negative contribution, resulting in a clearly positive splitting.  On the opposite side, MoSe$_2$ combines the smaller TM SOC and the largest CH SOC,  resulting in the largest negative contribution.  In MoS$_2$ the two competing contributions are the smallest (comparing to the other considered MLs) and go a long way to cancel each other: sulphur alone would give $\Delta\simeq -3$ meV whereas Mo alone would give $\Delta\simeq +2$ meV.

\section{Discussion and Conclusions}\label{Conclusions}
 
 We now discuss some of the limitations of our model.  First, it is apparent that the agreement between the TB+SOC model and the DFT+SOC results is not good in the case WSe$_2$. This is reflected in the discrepancy of the CB spin splitting shown in Table \ref{TAB1} and in the large variations of the  value of $\lambda_{\rm Se}$ determined using perturbation theory and the non-peturbative method (see Table \ref{tab_lambdas}).   This is due in part to the truncation in the number of bands in the TB method.  Interband contributions to bands omitted in the TB model contribute to the spin-orbit splitting, and this effect is of course larger for WSe$_2$ for which both $\lambda$'s are largest.   
 
 A second contribution to this discrepancy might arise from the fact that the MLWF are not exactly the same than atomic orbitals. However, the differences are large only in the interstitial region and should weakly affect the spin-orbit physics.  In contrast,  the loss of atomic symmetry in the interstitial region  clearly explains why our  attempts, not discussed above,  to parametrize the Wannier-TB Hamiltonian with a Slater Koster\cite{Slater-Koster} parameters have failed.  Therefore,  the method discussed in this paper needs to be modified in order to map the DFT calculation into a TB model parametrized with a few Slater Koster parameters, in the line of recent work.\cite{tight-binding1,tight-binding2,tight-binding3}  A third missing ingredient in the TB+SOC, compared to the DFT+SOC, are interatomic terms, as opposed to the intra-atomic contributions described in Eq. (\ref{ham_SOC}).
 
In summary,  DFT calculations show that  semiconducting two dimensional transition metal dichalcogenides have  spin orbit splittings at the conduction band that 
---although  smaller than those at the valence band--- are definitely large enough to be relevant experimentally. \cite{NatureComm2013-Delft}
In order to understand the chemical origin of the splitting, we have derived a tight-binding Hamiltonian 
(Eq. (\ref{ham_SOC2})) using  the maximally localized Wannier functions as a basis. Taking advantage of their atomic like character,  it is possible to add the atomic spin orbit coupling operators to the tight-binding model,  using the atomic $\lambda$ as adjustable parameters.   
 We have found that  this method describes very well the bands  in the energy range from -8 eV to 3 eV around
the Fermi level.  The tight-binding model permits to determine that both types of atoms, metal and chalcogen,
contribute to the conduction band spin splitting with opposite signs. This  naturally explains why conduction band spin-orbit
splittings of the WS$_2$ and MoSe$_2$ present opposite signs. 

Our findings have implications on a wide array of spin related  physical phenomena that are being explored in two dimensional transition metal dichalcogenides and their nanostructures,\cite{Klinovaja13}  including  the conduction band Landau Levels,\cite{Niu-LL-2013} spin relaxation, \cite{Dery-2013},  exciton spin selection rules,\cite{Jones-Nat-Nano13}  RKKY coupling, \cite{RKKY} as well as the spin and valley Hall effects.  \cite{Feng_PRB2012,Xiao,Shan13}

\section{Acnlowledgements}

We acknowledge fruitful discussions with  A. Korm\'anyos,  J. Jung,  A. H. MacDonald, I. Souza, J. L. Martins, E. V. Castro and J. L. Lado.  
We thankfully acknowledge the computer resources, technical expertise and assistance provided by the Red Espa\~nola de Supercomputaci\'{o}n.
 JFR acknowledges  financial supported by MEC-Spain (FIS2010-21883-C02-01) 
  and Generalitat Valenciana (ACOMP/2010/070), Prometeo. This work has been financially supported in part by FEDER funds.  We acknowledge financial support by Marie-Curie-ITN 607904-SPINOGRAPH. 

{\em Note Added}: Recently, related work has been posted (\onlinecite{Kormanys13b})

\end{document}